\documentclass[aip,jcp,amsmath,amssymb,longbibliography,reprint]{revtex4-1}

\usepackage{graphicx}
\usepackage{dcolumn}
\usepackage{bm}

\usepackage[utf8]{inputenc}
\usepackage[T1]{fontenc}
\usepackage{mathptmx}
\usepackage{etoolbox}
\usepackage{siunitx}
\usepackage{chemformula}
\usepackage{braket}
\DeclareSIUnit[number-unit-product = {\,}]
\cal{cal}

\makeatletter
\def\@email#1#2{%
 \endgroup
 \patchcmd{\titleblock@produce}
  {\frontmatter@RRAPformat}
  {\frontmatter@RRAPformat{\produce@RRAP{*#1\href{mailto:#2}{#2}}}\frontmatter@RRAPformat}
  {}{}
}%
\makeatother
\begin{document}


\title{Quantum criticality in chains of planar rotors with dipolar interactions}
\author{Tobias Serwatka}
\author{Pierre-Nicholas Roy}%
 \email{pnroy@uwaterloo.ca.}
\affiliation{ 
Department of Chemistry, University of Waterloo, Waterloo, Ontario N2L 3G1, Canada
}


\date{\today}

\begin{abstract}
In this contribution we perform a density matrix renormalization group study of chains of planar rotors interacting via dipolar interactions. By exploring the ground state from weakly to strongly interacting rotors, we find the occurrence of  a quantum phase transition between a disordered and a dipole-ordered quantum state. We show that the nature of the ordered state changes from ferroelectric to antiferroelectric when the relative orientation of the rotor planes varies and that this change requires no modification of the overall symmetry. The observed quantum phase transitions are characterized by critical exponents and central charges which reveal different universality classes ranging from that of the (1+1)D Ising model to the 2D classical XY model. 
\end{abstract}

\maketitle


\section{Introduction}

Dipolar many-body systems are very interesting as a platform for quantum devices as they provide a variety of phases with unique properties.\cite{carr2009cold,golomedov2011mesoscopic,samajdar2021quantum,baodipolar2023,connordipolar2023} 
At zero temperature, these systems show a special class of transitions between phases of different order, namely quantum phase transitions.\cite{vojta2003quantum,sachdev2011quantumcrit,sachdev2011quantum} 
Since there are no thermal fluctuations at zero temperature, these transitions are purely driven by quantum fluctuations.\cite{sachdev2000quantum,sachdev2011quantum} 
Experimentally, dipolar lattices are often realized and studied by confining dipolar diatomic molecules or atoms in excited states in optical traps.\cite{anderson2011trapping,yan2013observation} 
These optical lattices make it possible to control and tune the emerging phases with high precision. 
But they also require a huge effort to shield the lattice from the environment in order to avoid decomposition of the sensitive dipolar species and decoherence. An alternate route towards the realization of a dipolar lattice is to use chemically robust molecules with a permanent dipole moment such as water, and confine these species in chemical environments. 
In recent theoretical studies, we showed that one-dimensional systems of water or lithium fluoride confined in carbon nanotubes or fullerene cages can form ferroelectric quantum phases.~\cite{serwatka2022ferroelectric,serwatka2023quantum,serwatka2023endo} 
Path integral studies of general molecular rotors in one- and two-dimensional arrays show the occurrence of different quantum phases depending on the chosen lattice symmetry.~\cite{abolins2011ground,abolins2013erratum,abolins2018quantum}
Certain minerals such as beryl or cordierite also provide a confining environment.\cite{gorshunov2016incipient} 
These crystals have cavities in which water can be intercalated. The cavities are separated by distances of $\sim$\SI{5}{\angstrom} which suppresses the formation of hydrogen bonds between water molecules that only interact via their electric dipoles.\cite{belyanchikov2020dielectric} Experiments of water systems confined in cordierite crystals show that at temperatures below \SI{4}{\kelvin}, the water molecules start to align and ferroelectric phases are formed.\cite{belyanchikov2020dielectric} 
Interestingly, studies of beryl and cordierite systems suggest that in these crystals water does not behave as a general $O(3)$ rotor but the dipole of water preferentially moves in a plane.~\cite{kolesnikov2016quantum,belyanchikov2022single} That is due to the potential of six-fold symmetry provided by the cavity which causes a quantum tunneling of water~\cite{kolesnikov2016quantum}. Hence, the whole system can be simplified as an assembly of planar (or $O(2)$) rotors interacting which each other via dipolar interactions.

Different computational approaches can be used to compute the quantum states of rotating molecules with dipolar couplings. 
Exact diagonalization (ED) techniques can be applied to dimers\cite{felker2017electric} and assemblies containing up to six molecules.\cite{felker2017accurate}
Systems containing up to ten molecules can be computed by ED using basis truncation schemes.\cite{halverson2018quantifying}
For larger system sizes, paths integral techniques can be used to obtain energetic and structural properties,\cite{abolins2011ground,abolins2013erratum,abolins2018quantum} and R\'enyi entanglement
via the so-called replica trick.\cite{sahoo2020path,sahoo2023effect} 
Path integral approaches also allow the treatment of asymmetric top dipolar rotors.\cite{sahoo2021path}
The density matrix renormalization group (DMRG) method\cite{white1992density} is however the most practical choice for large one-dimensional lattices of interacting dipolar rotors.\cite{iouchtchenko2018ground,mainali2021comparison,serwatka2022ground,serwatka2022ferroelectric,serwatka2023quantum}

In this study, we  focus on different linear chains of planar dipolar rotors, ranging from coplanar to stacked arrangements. In Sec.~\ref{sec:theory} we  present the system's Hamiltonian and  briefly motivate and explain the DMRG method used. This is followed by computational details given in Sec.~\ref{sec:comp}. In Sec.~\ref{sec:results}, we present and discuss the different quantum phases and characterize the phase transition. 
We conclude and provide a perspective in Sec.~\ref{sec:conclusio}.      

\section{Methodology}\label{sec:theory}

\subsection{Planar rotor chains}

\begin{figure}
\includegraphics[width=\columnwidth]{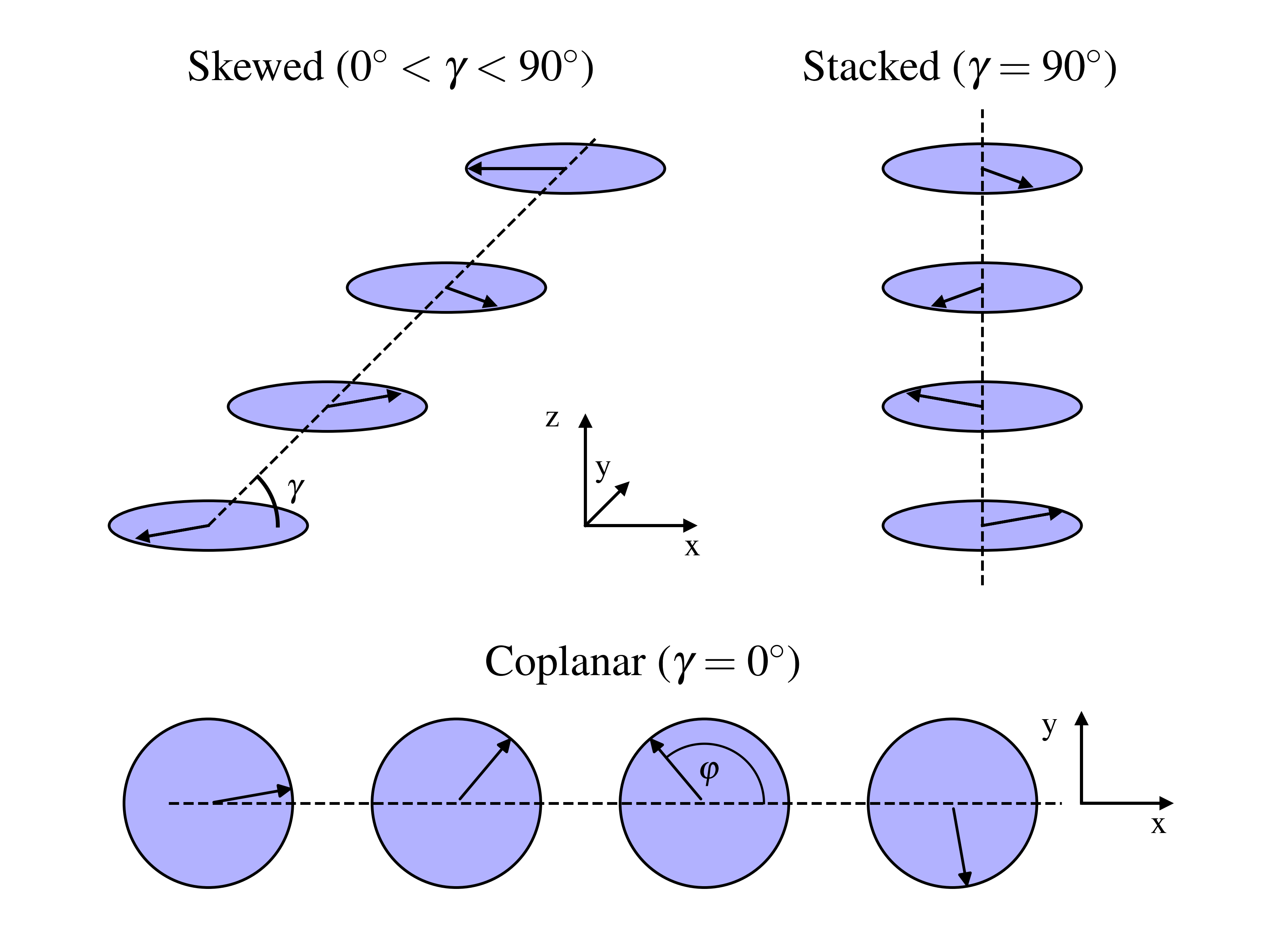}
\caption{Sketch of the relative arrangements the dipolar planes can have depending on the chain angle $\gamma$.}
\label{fig:system}
\end{figure}

We consider chains of planar or $O(2)$ quantum rotors interacting via dipole-dipole interactions. In the case of $N$ rotors with nearest-neighbor interactions, these systems are described by the following Hamiltonian,
\begin{align}
\hat{H} = \sum_{i=1}^{N}\hat{L}^2_{z,i}+g\sum_{i=1}^{N-1}\left[\hat{e}_{i}\hat{e}_{i+1}-3(\hat{e}_{i}\hat{r})(\hat{e}_{i+1}\hat{r})\right],
\label{eq:Ham1}
\end{align}
where the first sum specifies the rotational kinetic energy that depends on $\hat{L}_{z,i}$, the angular momentum of site $i$ perpendicular to the rotor plane. The second sum denotes the dipolar interaction term whose strength is determined by $g$. The interaction between a pair of rotors depends on the dipole orientation vector $\hat{e}=(\hat{x},\hat{y})^{T}=(\cos\varphi,\sin\varphi)^{T}$ and the unit vector $\hat{r}$ connecting the two sites. The orientation of the dipole vector can be parametrized by the angle $\varphi$ as illustrated in Fig.~\ref{fig:system}. By choosing an appropriate frame for the whole chain, the Hamiltonian in Eq.~\eqref{eq:Ham1} can be written as
\begin{align}
\hat{H}^{\rm CF} = \sum_{i=1}^{N}\hat{L}^2_{z,i}+g\sum_{i=1}^{N-1}\left[\hat{y}_{i}\hat{y}_{i+1}+(1-3\cos^2(\gamma))\hat{x}_{i}\hat{x}_{i+1}\right],
\label{eq:Ham2}
\end{align}
where $\gamma\in (0^\circ,90^\circ)$ denotes the angle between $\hat{r}$ and the rotor planes and the superscript CF stands for chain frame (see Fig.~\ref{fig:system}).
Modulating $\gamma$ allows one to go from a coplanar chain $\gamma=0^\circ$
\begin{align}
  \hat{H}^{\rm CF} = \sum_{i=1}^{N}\hat{L}^2_{z,i}+g\sum_{i=1}^{N-1}\left[\hat{y}_{i}\hat{y}_{i+1}-2\hat{x}_{i}\hat{x}_{i+1}\right],
\label{eq:Ham3}
\end{align}
to a chain of stacked rotors ($\gamma=90^\circ$)
\begin{align}
  \hat{H}^{\rm CF} = \sum_{i=1}^{N}\hat{L}^2_{z,i}+g\sum_{i=1}^{N-1}\left[\hat{y}_{i}\hat{y}_{i+1}+\hat{x}_{i}\hat{x}_{i+1}\right].
\label{eq:Ham4}
\end{align}
The latter Hamiltonian is the well known quantum rotor chain.~\cite{sachdev2011quantum}
A third special configuration exists at $\gamma=\arccos\left(\sqrt{2/3}\right)$, at which the Hamitonian becomes
\begin{align}
  \hat{H}^{\rm CF} = \sum_{i=1}^{N}\hat{L}^2_{z,i}+g\sum_{i=1}^{N-1}\left[\hat{y}_{i}\hat{y}_{i+1}-\hat{x}_{i}\hat{x}_{i+1}\right].
\label{eq:Ham5}
\end{align}
where the sign of the $x$-term goes from positive to negative compared to Eq. \eqref{eq:Ham4} (see Appendix \ref{sec:appendix} for a symmetry analysis of the interaction).

\subsection{Density matrix renormalization group}

In order to represent the ground states of the aforementioned Hamiltonians, the wavefunctions are expanded in a tensor product basis,
\begin{align}
\ket{\Psi}=\sum_{m_{1},...,m_{N}}C_{m_{1}\cdots m_{N}}\Ket{m_{1}}\otimes\cdots\otimes\Ket{m_{N}},
\label{eq:wf}
\end{align}
where the $\hat{L}_{z}$ eigenbasis $\lbrace\Ket{m}\rbrace_{m=-m_{\mathrm{max}},...,m_{\mathrm{max}}}$ is used as the local basis set. Instead of working with the full expansion tensor in Eq.~\eqref{eq:wf} it is decomposed as a matrix product state (MPS),
\begin{align}
C_{m_{1}\cdots m_{N}}&=\sum_{\alpha_{1},...,\alpha_{N-1}}A_{1\alpha_{1}}^{m_{1}}\cdots A_{\alpha_{N-1}1}^{m_{N}}\\
&= \mathbf{A}^{m_{1}}\cdots\mathbf{A}^{m_{N}}.
\label{eq:MPS}
\end{align} 
 Open-boundary conditions are employed in Eq.~\eqref{eq:MPS}. The dimensions of the matrices $\mathbf{A}^{m}$ are termed bond dimensions. For many systems these bond dimensions can be chosen much smaller than algebraically required by Eq.~\eqref{eq:MPS} which leads to a still accurate but much more compact representation.\cite{schollwock2011density} In order to optimize the matrices in Eq.~\eqref{eq:MPS}, the DMRG approach is employed~\cite{white1992density}. A central property in the DMRG optimization is the von-Neumann entanglement entropy,
\begin{align}
S_{\mathrm{vN}} = -\mathrm{tr}\left(\hat{\rho}_{A}\ln\left(\hat{\rho}_{A}\right)\right)
\end{align}
where $\hat{\rho}_{A}$ is the reduced density operator of a subregion $A$ of the whole system. The von-Neumann entropy provides a measure of the entanglement between subregion $A$ and the remaining subregion $B$. In the DMRG optimization, the boundary of $A$ is swept from one end of the chain to the other. During these sweeps, iterative diagonalization and singular value decomposition (SVD) are used to optimize matrices with bond dimension small enough for a compact representation, but large enough to describe the entanglement entropy between $A$ and $B$ faithfully. For a more detailed description of the method see Refs.~\onlinecite{schollwock2011density,serwatka2022ground}.


\section{Computational details}\label{sec:comp}
In order to optimize the MPS representation of the ground states, we employ a two-site DMRG algorithm as implemented in ITensor.\cite{fishman2022itensor} An SVD threshold of $10^{-8}$ and a relative energy cut-off of $10^{-8}$ are used. The local Hilbert spaces are described by a plane wave eigenbasis with $m_{\mathrm{max}}=4$. For the calculations of central charges and critical exponents we modify the Hamiltonian by a sine-squared deformation\cite{katsura2012sine,hotta2012grand} (SSD) in order to mimic periodic boundary conditions in open-boundary simulations.

\section{Results and Discussion}\label{sec:results}

\subsection{Quantum phases}

\begin{figure}
\includegraphics[width=\columnwidth]{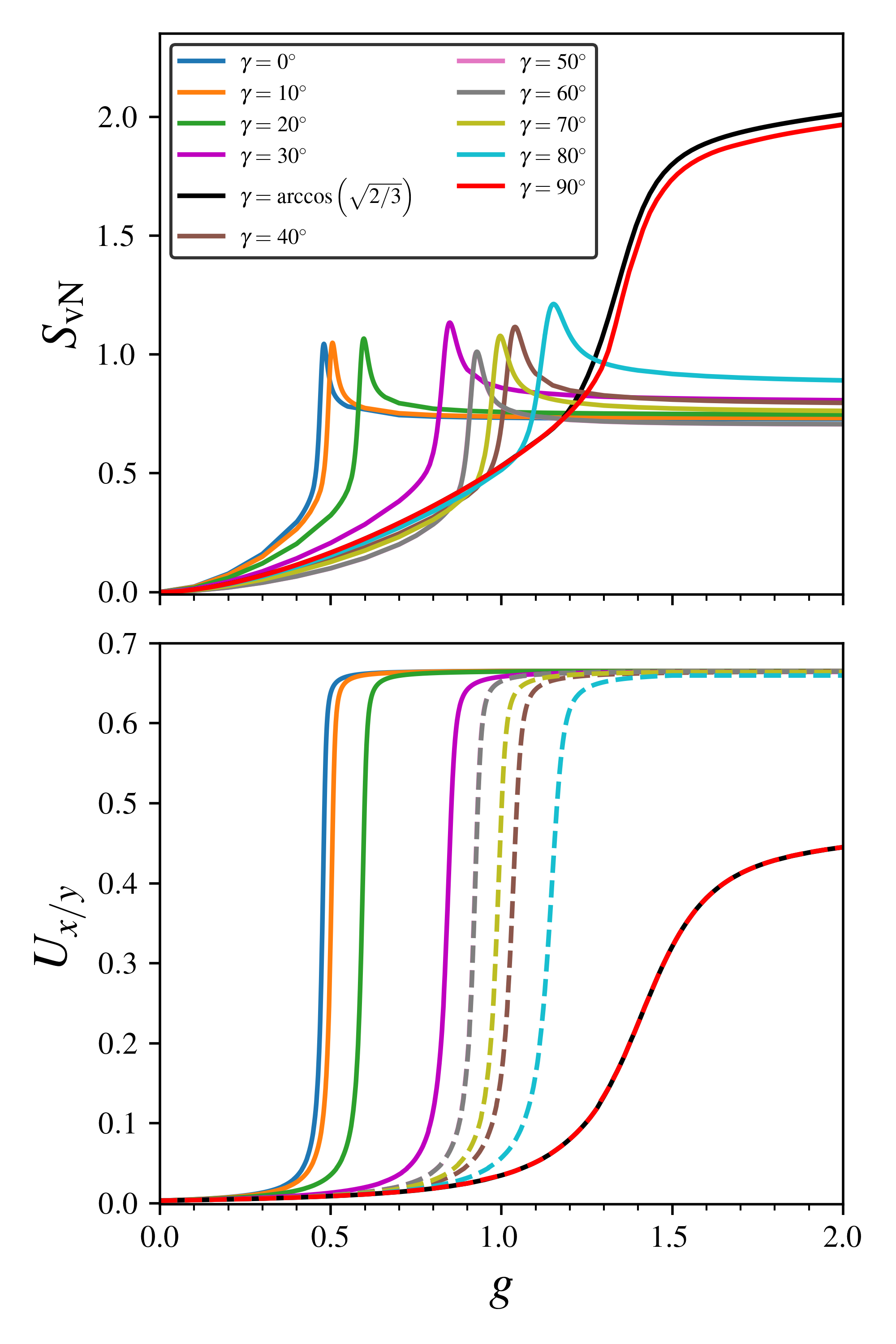}
\caption{von-Neumann entanglement entropy (top panel) for different chain angles $\gamma$ and interaction strengths $g$. Binder parameter $U_{a}=1-\frac{\langle M_{a}^4\rangle}{3\langle M_{a}^2\rangle^2}$ $(a=x,y)$ (bottom panel) for different chain angles $\gamma$ at $g=2$. For $\gamma<\arccos\left(\sqrt{2/3}\right)$ the polarization $M_{x}=\sum_{i=1}^{N}x_{i}$ is employed to define the Binder parameter (dashed lines). For $\gamma\geq \arccos\left(\sqrt{2/3}\right)$ the staggered polarization $M_{y}=\sum_{i=1}^{N}(-1)^{i}y_{i}$ is employed to define the Binder parameter (solid lines). All entropy calculations are performed for chains with $N=50$ (with SSD) and all Binder parameter calculations are performed with $N=150$.}
\label{fig:entropy_binder}
\end{figure}

\begin{figure*}
\includegraphics[width=\textwidth]{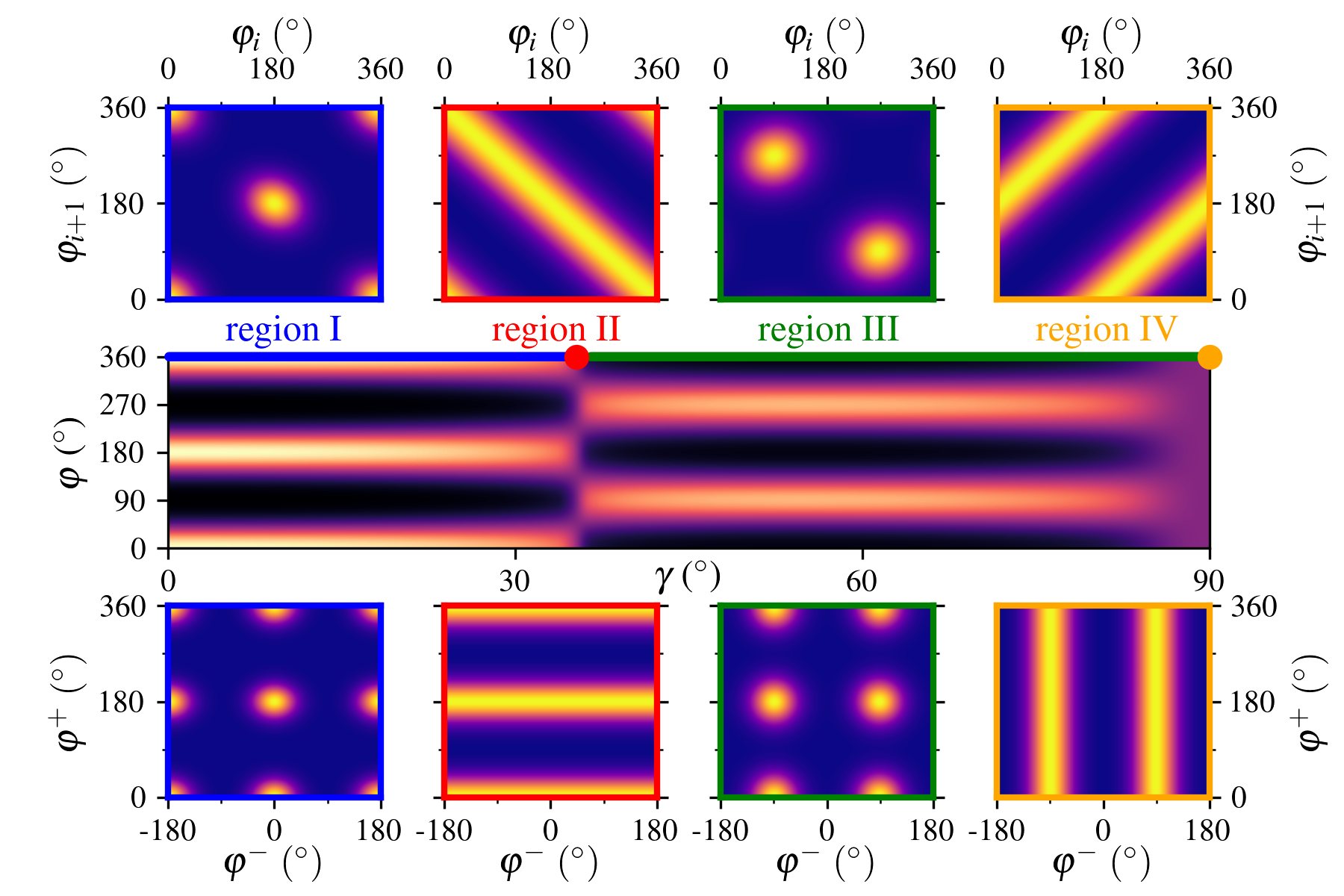}
\caption{Pair density $P(\varphi_i,\varphi_{i+1})$ of the central sites for $\gamma=0^\circ, \arccos\left(\sqrt{2/3}\right), 60^\circ, 90^\circ$ (top panels). One-site density $P(\varphi)$ at different $\gamma$ of one of the central sites (middle panels). Pair density $P(\varphi^{+},\varphi^{-})$ of the central sites for $\gamma=0^\circ, \arccos\left(\sqrt{2/3}\right), 60^\circ, 90^\circ$ (bottom panels). All calculations are performed for a chain with $N=10$ at $g=2.0$.}
\label{fig:density_angles}
\end{figure*}

\begin{figure*}
\includegraphics[width=\textwidth]{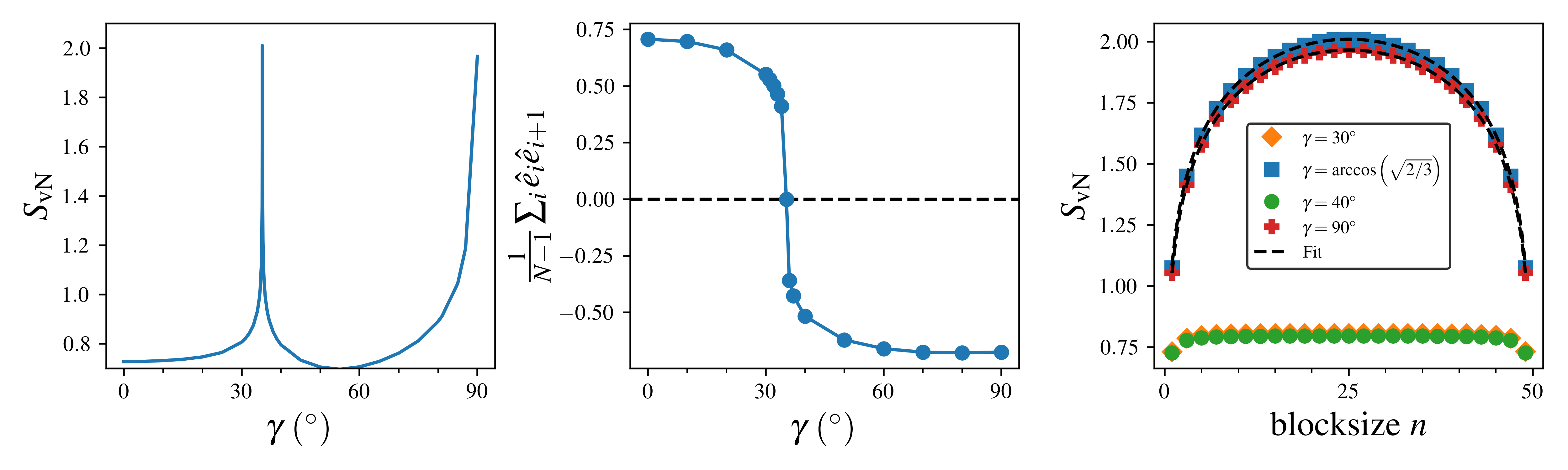}
\caption{von-Neumann entanglement entropy for different chain angles $\gamma$ (left panel). Nearest-neighbor correlation function for different chain angles $\gamma$ (central panel). von-Neumann entanglement entropy for different partitions of the chain calculated in the critical region ($\gamma=\arccos\left(\sqrt{2/3}\right)$ and $90^\circ$) and away from it ($\gamma=30^\circ$ and $40^\circ$) (right panel). The curve in the critical region was fitted to $c/3\ln\left[N/\pi\sin\left(n\pi/N\right)\right]$ with $c=1.0029\pm 0.0002$ for $\gamma=35.3^\circ$ and $c=0.9883 \pm 0.0003$ for $\gamma=90^\circ$. All calculations are performed for a chain with $N=50$ (with SSD) at $g=2.0$.}
\label{fig:area_angles}
\end{figure*}

\begin{figure}
\includegraphics[width=\columnwidth]{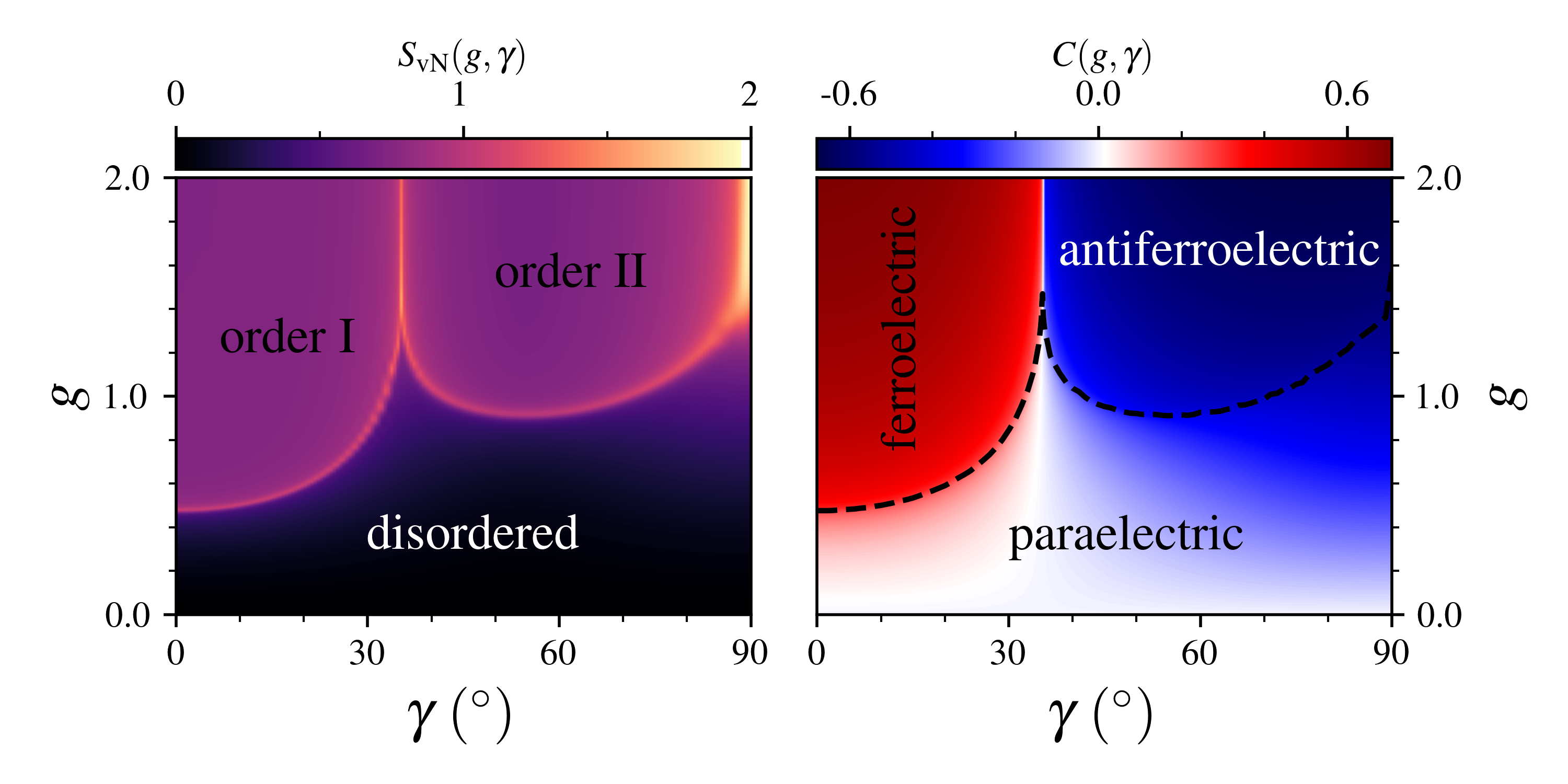}
\caption{The $g$- and $\gamma$-dependence of the von-Neumann entanglement entropy and the associated quantum phases (left panel). The $g$- and $\gamma$-dependence of the nearest-neughbor correlation function $C=\frac{1}{N-1}\sum_{i=1}^{N-1}\hat{e}_{i}\hat{e}_{i+1}$ and the associated quantum phases (right panel). The dotted line denotes the maximum of the derivative of the correlation function, $\mathrm{max}\left(\frac{\partial}{\partial g}C\right)_{\gamma}$. For all fits, a chain with $N=50$ (with SSD) was employed. }
\label{fig:phasediagram}
\end{figure}

In a first step we want to study if and for which parameter values $g$ and $\gamma$ quantum phase transitions occur in the planar rotor chains. For this purpose, we calculate the bipartite von-Neumann entanglement entropy $S_{\mathrm{vN}}$ as this quantity is known to be a good indicator for such transitions.\cite{serwatka2022ground,serwatka2023quantum} We performed calculations for different chain angles $\gamma$ and scanned the entanglement entropy along $g$ (see Fig.~\ref{fig:entropy_binder}). In these calculations a SSD was employed because it decreases boundary effects.\cite{katsura2012sine}
In Fig.~\ref{fig:entropy_binder}, one can observe that for almost all chain angles $\gamma$, the entanglement entropy increases with $g$ up to a maximum followed by a plateau. The peak marks the transition from a disordered to a dipole-ordered quantum phase. There are only two angles for which a different behavior is observed. At $\gamma=\arccos\left(\sqrt{2/3}\right)$ and $\gamma=90^\circ$ there is no maximum, the entanglement entropy keeps rising but with a smaller rate after $g\approx 1.4$. This different behavior is due to a change in symmetry of these chains. For all $\gamma$ values except these two, the chains obey a $\mathbb{Z}_{2}$ symmetry, i.e. the system is invariant with respect to a collective rotation of all dipoles by $180^\circ$. Hence, with $\mathbb{Z}_{2}$ symmetry, the ground state in the ordered phase is two-fold degenerate ({\em near} degenerate for finite $N$) and can be expressed as a superposition of a left-and right-polarized state. Only at the quantum phase transition the Hamiltonian becomes ungapped and a superposition of infinitely many equally weighted states occurs which leads to the singularity peak in $S_{\mathrm{vN}}$. For $\gamma=\arccos\left(\sqrt{2/3}\right)$ and $\gamma=90^\circ$ the symmetry group changes from the discrete $\mathbb{Z}_{2}$ to the continuous $U(1)$, where the system is invariant with respect to a rotation by an arbitrary angle. The difference between $\gamma=\arccos\left(\sqrt{2/3}\right)$ and $\gamma=90^\circ$ is that in the second case, all dipoles are rotated by the same angle $\Delta\varphi\in (0^\circ,360^\circ)$ while for the first case, neighboring dipoles are rotated by an angle $\Delta\varphi$ and $360^\circ-\Delta\varphi$ with $\Delta\varphi\in (0^\circ,360^\circ)$ (see Appendix \ref{sec:appendix}). Thus, due to the continuous symmetry even away from the quantum phase transition in the ordered phase, there is an equally-weighted superposition of an infinite number of states and no peak in $S_{\mathrm{vN}}$ will appear. In Fig.~\ref{fig:entropy_binder} we also depict the Binder parameter which appears to be a good order parameter, i.e. a property that is zero in the disordered phase and has some finite value in the ordered phase.\cite{binder1981finite,binder1981critical} 
Here we define the Binder parameter with respect to the polarization operator $M_{a}$ $(a=x,y)$. For chain angles $\gamma<\arccos\left(\sqrt{2/3}\right)$, the polarization operator along $x$ and for larger angles the staggered polarization operator along $y$ are used. As we will see later, that change in operator is required by the change in the ordered quantum phase.
Both the entropy and the Binder curves in Fig.~\ref{fig:entropy_binder} not only show that a transition occurs but also for which interaction strength $g_{c}$. For increasing chain angles $\gamma$, this critical strength $g_{c}$ shifts to larger values up to $\gamma=\arccos\left(\sqrt{2/3}\right)$ after which it falls back to smaller values and starts to rise again until $90^\circ$. This shift is due to the competition of kinetic and potential terms in Eq.~\eqref{eq:Ham2}. The kinetic terms are the same for all $\gamma$ whereas the potential depends on $\gamma$. For a given $g$ and $\gamma=0^\circ$, the lowest potential energy is obtained by a parallel alignment of all dipoles. By increasing $\gamma$ the potential energy for alignment gets smaller. To compensate for that the interaction has to be stronger thus $g_{c}$ shifts to larger values. From $\gamma=\arccos\left(\sqrt{2/3}\right)$ on, it becomes energetically more favorable to have antiparallel dipoles. This structural change leads at first to a fall of $g_{c}$ which starts to rise. Again, that increase is due to a less favorable alignment of the dipoles.\\
Since we could show the occurrence quantum phase transitions and thus ordered quantum phases, we want to explore the nature of these phases. To do so, we examine the angular density $P(\varphi)$ of a central site shown in the center panel of Fig.~\ref{fig:density_angles}. 
For every chain angle $\gamma$, the density is calculated at $g=2.0$ to ensure that the systems is deep in the ordered phase. 
The one-site density shown in the middle panels of Fig.~\ref{fig:density_angles} reveals four distinct regions (I-IV) characterized by different localization patterns. In region I with $0^\circ\leq\gamma<\arccos\left(\sqrt{2/3}\right)$, the rotor is localized around $\varphi=0^\circ$ and $\varphi=180^\circ$, i.e. parallel and antiparallel to the chain axis. Region II can be identified at $\gamma=\arccos\left(\sqrt{2/3}\right)$ for which the rotor is completely delocalized along $\varphi$. That region is followed by region III for $\arccos\left(\sqrt{2/3}\right)<\gamma< 90^\circ$. The density pattern is similar to that of region I but shifted by $90^\circ$, i.e. the rotor is localized around $\varphi=90^\circ$ and $\varphi=270^\circ$ (perpendicular to the chain axis). Region IV occurs at $\gamma=90^\circ$ and resembles region II as it is delocalized over $\varphi$. In order to elucidate the nature of the ordered phases we went one step further and calculated the (joint) pair density $P(\varphi_{i},\varphi_{i+1})$ of the two central sites at $\gamma=0^\circ, \arccos\left(\sqrt{2/3}\right), 60^\circ$ and $90^\circ$ (see upper panels of Fig.~\ref{fig:density_angles}). This allows us to gain insight into the relative orientation of neighboring sites. The pair densities of region I and III reveal clearly that the ordered phases in both regions are of a different kind. The rotors in both regions are not only rotated as shown by $P(\varphi)$ but in region I, all dipoles point in the same direction whereas in region III, neighboring dipoles point in opposite directions. Hence, considering spontaneous symmetry breaking, region I is characterized by a ferroelectric quantum phase whereas region III has an antiferroelectric phase. The pair density is also very useful to characterize the transition regions II and IV because based on the one-site density $P(\varphi)$, these regions cannot be distinguished from an disordered phase. By looking at the joint $P(\varphi_{i},\varphi_{i+1})$, the localization reveals that for all regions, the system is in an ordered quantum phase. The delocalization along certain axes in the pair densities for region II and IV is due to the $U(1)$ symmetry occurring at the special angles associated with both regions. To make the relative orientation more evident, it is useful to transform the pair density to the relative coordinates $\varphi^{+}=\frac{1}{2}(\varphi_{i}+\varphi_{i+1})$ and $\varphi^{-}=\frac{1}{2}(\varphi_{i}-\varphi_{i+1})$ as shown in the bottom panels of Fig.~\ref{fig:density_angles}. The relative orientation of the dipoles which determines the nature of the quantum phases is nicely highlighted by the pair density along $\varphi^{-}$. For the antiferroelectric systems (region III and IV), the densities are localized around $\varphi^{-}=\pm 90^\circ$ and for the ferroelectric quantum phase (region I), they are localized around $\varphi^{+}=n\cdot 180^\circ, n=0,\pm 1$. 
The delocalization in one of the coordinates $\varphi^{+}$ or $\varphi^{-}$ reflects the $U(1)$ symmetry character. The regions with $\mathbb{Z}_{2}$ symmetry are strongly localized along both axes. 
As Fig.~\ref{fig:density_angles} shows, region II is the transition between the ferroelectric region I and the antiferroelectric region III. Thus, its ground state is a superposition of two ferro- and two antiferroelectric states and an infinite number of ferrielectric states due to the continuous $U(1)$ symmetry. As required by the special symmetry of $\hat{H}$ in that region (see Appendix \ref{sec:appendix}), these states have to obey $\varphi_{1}+\varphi_{2}=n\cdot 360^\circ,n=0,\pm 360^\circ$ which can be seen clearly in the pair density pattern of region II in Fig.~\ref{fig:density_angles}. Hence, in contrast to the other three regions, the transition region II is not of a pure but rather of a mixed character.\\
It is interesting to observe that systems with the same overall symmetry (and thus the same critical behavior as we will show later) have structurally different ground states which depend on the relative orientation of the rotor planes within the chains. In fact, this change from parallel to antiparallel alignment constitutes a second quantum phase transition in this system. But this time it is not the kind of transition we found earlier, which occurs between disordered and ordered phases. This second kind of transition marks the passage between distinct ordered phases. Earlier, in Fig.~\ref{fig:entropy_binder}, we showed that this first kind of transition is signalled by a peak in the entanglement entropy (for $\mathbb{Z}_{2}$ symmetry) and an inflection point in the Binder parameter when these properties are scanned along $g$ for a fixed $\gamma$. A similar strategy can be employed to spot the second phase transition. Now, the parameters are reversed and $g$ is fixed while the chain angle $\gamma$ is scanned. An appropriate parameter to signal the transition is the nearest-neighbor correlation function $C$ depicted in Fig.~\ref{fig:area_angles}. We calculated it at $g=2.0$. The correlation function changes its sign when going from the ferroelectric phase in region I to the antiferroelectric phase in region III. For region II it is zero because all negative and positive contributions of the superposition of states cancel out. In contrast to the pair densities the correlation function is not able to detect the transition between the two antiferroelectric phase in region II and III. So we also calculated the entanglement entropy. The resulting curve is depicted in Fig.~\ref{fig:area_angles}. 
It shows a peak around $\gamma=\arccos\left(\sqrt{2/3}\right)$ corresponding to the transition from a ferro- to an antiferroeletric phase already observed in the angular density of Fig.~\ref{fig:density_angles}. At $\gamma=90^\circ$ there is a second peak in the entropy curve in Fig.~\ref{fig:area_angles} which is due to the change in symmetry from $\mathbb{Z}_{2}$ to $U(1)$. This change does not affect the relative orientation within the quantum phase, but it increases the degeneracy of the ground state. We observe that both kinds of quantum phase transitions, disorder-order and order-order can be detected by peaks in the entanglement entropy. 
In the first kind, the entropy curves, depicted in the top panel of Fig.~\ref{fig:entropy_binder}, show a pronounced step because the systems goes from a disentangled or weakly entangled, to a strongly entangled phase. 
But for the second kind, the entanglement entropy, depicted in Fig.~\ref{fig:area_angles} has similar values on both sites of the peak because only the specific order changes, not the entanglement. 
We can connect these findings to the pair density in the relative coordinates shown in Fig.~\ref{fig:density_angles}. 
The disorder-order transition, and the entropy jump are associated with a change of the localization pattern of $P(\varphi^{+},\varphi^{-})$ in both relative coordinates $\varphi^{+}$ and $\varphi^{-}$ because for a disordered phase $P(\varphi^{+},\varphi^{-})=const.$. In contrast, the order-order transition is associated with a pair density change in only one of the relative coordinates $\varphi^{+}$ or $\varphi^{-}$. The ferroelectric-antiferroelectric transition changes the pair density along $\varphi^{-}$ whereas the $\mathbb{Z}_{2}$-$U(1)$ transition changes the density along $\varphi^{+}$. \\
The different quantum phases and transitions we found and their $(g,\gamma)$ dependence are summarized in two phase diagrams depicted in Fig.~\ref{fig:phasediagram}.
The phase diagram based on the entanglement entropy highlights the transitions from disordered to ordered, and between ordered phases. However, it does not reveal the precise nature of the order I and order II quantum phases. This information is provided by the correlation function which allows a clear distinction between para-, ferro- and antiferroelectric quantum phases.

\subsection{Critical exponents - Universality class}

\begin{figure}
\includegraphics[width=\columnwidth]{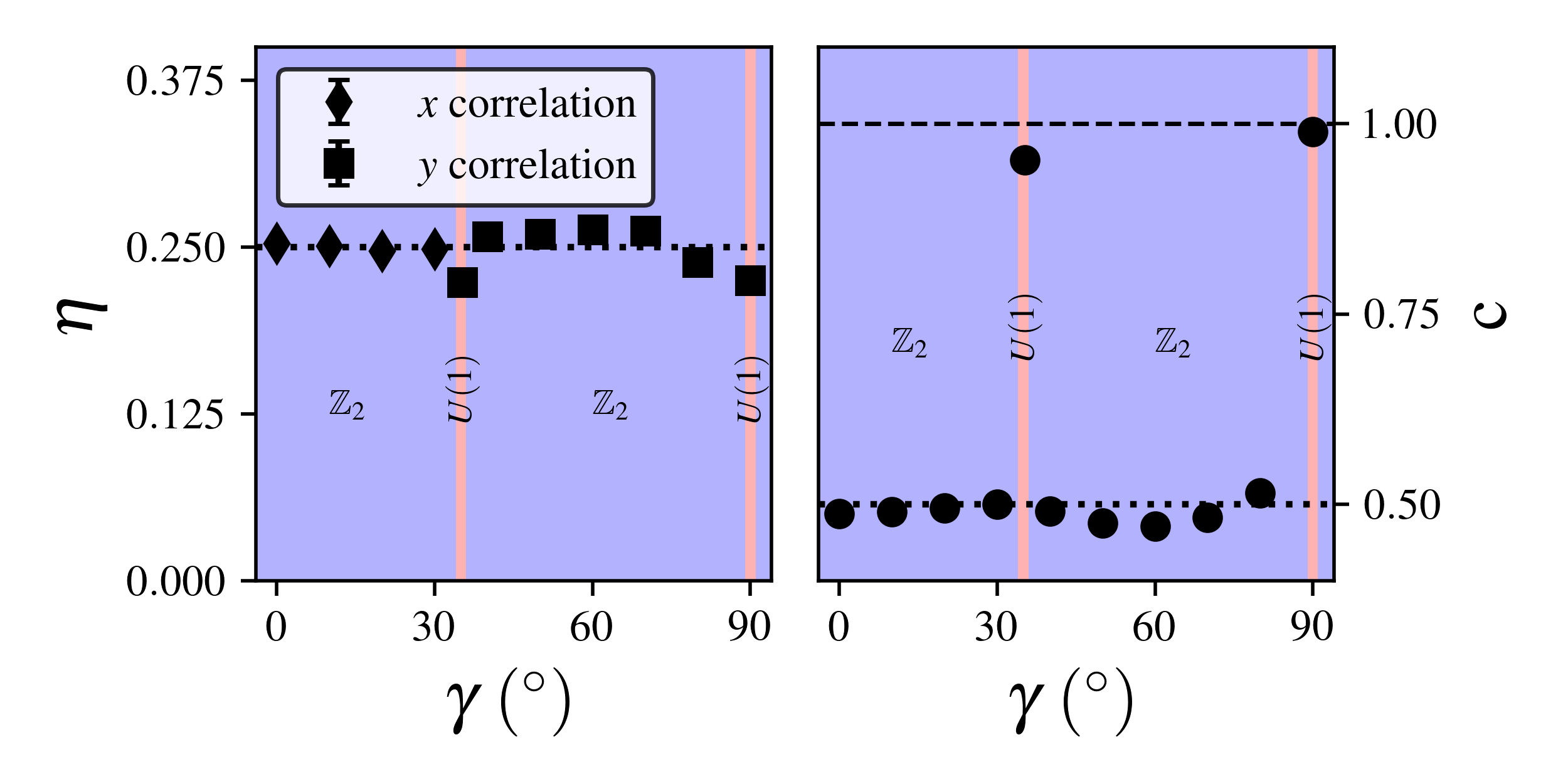}
\caption{Critical exponent $\eta$ obtained for different angles $\gamma$ (left panel). The exponents are determined by fitting the $x$ or $y$ correlation function of the central site and its neighbors at the critical interaction strength to the powerlaw $r^{-\eta}$. Central charge $c$ obtained for different angles $\gamma$ (right panel). The central charges are determined by fitting the von-Neumann entanglement entropy for different partitions $n$ at the critical interaction strength to $c/3\ln\left[N/\pi\sin\left(n\pi/N\right)\right]$. The different symmetries are marked by shaded areas. For all fits a chain with  $N=151$ (and with SSD) was employed.}
\label{fig:crit_angles}
\end{figure}

Having characterized the different ordered phases, we now focus on the quantum phase transitions in more details. For this, we calculate various properties in the critical region and fit them to analytic expressions. In that way, it is possible to obtain critical exponents and central charges which determine the universality class of the quantum phase transitions.\cite{calabrese2009entanglement,sachdev2011quantum,serwatka2023quantum} 
We start with the disorder-order phase transitions at various chain angles $\gamma$. First we look at the dipole correlation function of the central site $m$ and its neighbors in a chain with odd $N$. Away from criticality, this correlation function decays exponentially but at the transition, where the entanglement diverges for most angles, the correlations decay with a power law $r_{ij}^{-\eta}$ where $r_{ij}$ labels the distance between site $i$ and $j$.\cite{sachdev2011quantum} By fitting that expression to calculated pair correlation curves we determine the critical exponent $\eta$. The calculated exponents for various chain angles $\gamma$ are depicted in Fig.~\ref{fig:crit_angles}.
We used the $\langle x_{m}x_{m+i}\rangle$ correlation functions for all $\gamma<\arccos\left(\sqrt{2/3}\right)$ and the $(-1)^{i}\langle y_{m}y_{m+i}\rangle$ staggered correlation functions for all $\gamma\geq \arccos\left(\sqrt{2/3}\right)$. Fig.~\ref{fig:crit_angles} shows that a critical exponent of about 0.25 is obtained for all chain angles no matter if the system obeys $\mathbb{Z}_{2}$ or $U(1)$ symmetry. To illustrate that both groups belong to different universality classes, we calculate another kind of coefficient, namely the central charge $c$.\cite{calabrese2004entanglement,calabrese2009entanglement} This central charge can be obtained by calculating the entanglement entropy for different partitions of the chain at the transition. In gapped systems away from criticality, these entropy curves obey the area law, i.e. in one-dimensional systems they are constant.\cite{eisert2010colloquium} 
But at the transition, where the energy gap closes, the area law is violated and the entropy scales logarithmically. By fitting the entropy curves to the analytic expression $\left(c/3\right)\ln\left[(N/\pi)\sin\left(n\pi/N\right)\right]$ derived from conformal field theory, the central charge $c$ can be determined.\cite{calabrese2004entanglement} 
When we look at the calculated values depicted in Fig.~\ref{fig:area_angles}, we see that for all chain angles except $\gamma=\arccos\left(\sqrt{2/3}\right)$ and $\gamma= 90^\circ$, the central charge is about 0.5 which corresponds to the value for the (1+1)D Ising universality class.\cite{ibarra2016hobbyhorse} At $\gamma=\arccos\left(\sqrt{2/3}\right)$ and $\gamma= 90^\circ$, the symmetry changes from $\mathbb{Z}_{2}$ to $U(1)$ and this time, the central charge changes too. It is 1 which corresponds to the value for the 2D classical XY universality class.\cite{kosterlitz1974critical} 
These two different central charges also match the calculated $\eta$ values as both universality classes have the same value for $\eta$.\cite{sachdev2011quantum} 
For the critical exponent $\eta$, the values for $\gamma=\arccos\left(\sqrt{2/3}\right)$ and $\gamma= 90^\circ$ deviate more strongly from the analytic values than the values for the other angles. These differences between calculated and analytic values are only due to finite-size effects and numerical limitations. 
In particular, $\eta$ appears to be very sensitive and requires large $N$ and very fine grids to detect the critical $g_{c}$. When approaching $\gamma=\arccos\left(\sqrt{2/3}\right)$ or $\gamma=90^\circ$, the computational cost increases further due to the increasing entanglement entropy (see left panel in Fig.~\ref{fig:area_angles}) which leads to less accurate values for $\eta$. To reduce the numerical effort for the multitude of required alculations, we limit the system size to $N=151$ which gives a reasonable estimate of $\eta$ over all $\gamma$ considered in this study. The right panel of Fig.~\ref{fig:area_angles} shows that for most angles, such as $\gamma=30^\circ$ or $40^\circ$ away from the quantum phase transition the area law is obeyed and the entropy is a constant. Interestingly, for $\gamma=\arccos\left(\sqrt{2/3}\right)$ and $\gamma=90^\circ$ one observes the logarithmic behavior of $S_{\mathrm{vN}}$, hence the violation of area law not only in the critical region but also deep in the ordered phase. By fitting these to the analytic expression, again one obtains a central of 1. The logarithmic behavior is due to the infinite degeneracy of the ground state at $U(1)$ symmetry. Hence, there is a vanishing gap within the ground state manifold which leads to increasing $S_{\mathrm{vN}}$ even in the ordered phase as discussed earlier.

\section{Concluding Remarks}\label{sec:conclusio}
 
The quantum phases of dipolar rotor chains have been investigated using the DMRG approach. For every chain angle $\gamma$, we find that by increasing the interaction strength $g$, a quantum phase transition from a disordered to a dipole-ordered quantum phase occurs. By calculating critical exponents as well as central charges, we  characterized the phase transitions. For most chain angles $\gamma$, due to the $\mathbb{Z}_{2}$ symmetry, the corresponding quantum phase transitions belong to the (1+1)D Ising universality class. Only at two angles, $\gamma=\arccos\left(\sqrt{2/3}\right)$ and $90^\circ$, the system's symmetry changes to $U(1)$ which leads to a 2D classical XY universality class. For these two angles, logarithmic behavior of the entanglement entropy is found even for any $g\geq g_{c}$.\\
Apart from the quantum phase transitions, we also studied the ordered phases in more detail. Depending on the chain angle $\gamma$, the ordered phase can have a ferroelectric or an antiferroelectric ground state for the chains with $\mathbb{Z}_{2}$ symmetry. These two regions are separated by a phase with $U(1)$ symmetry at $\gamma=\arccos\left(\sqrt{2/3}\right)$. This quantum phase is of a mixed character as it is a superposition of ferro-, antiferro-and ferrielectric states. It could be pictured as a chain of counter rotating dipoles.\\
The occurrence of quantum phases of different kinds is of interest for chemical systems such as crystals because it shows that it might be possible to realize completely different phases (ferroelectric, antiferroelectric) by manipulating the crystal structure without changing the overall symmetry. For future work, it will be interesting to study if and how the quantum phase transition and the ordered phases are affected by introducing local symmetries such as local potentials with $n$-fold wells as found in beryl crystals.\cite{gorshunov2016incipient,kolesnikov2016quantum} 

\section*{Acknowledgements}
P.-N. R. acknowledges the Natural Sciences and Engineering Research Council (NSERC) of Canada, the Ontario Ministry of Research and Innovation (MRI), the Canada Research Chair program (950-231024), and the Canada Foundation for Innovation (CFI). T. S. acknowledges a Walter-Benjamin funding of the Deutsche Forschungsgemeinschaft (Projektnummer 503971734). 

\section*{DATA AVAILABILITY}
The data that support the findings of this study are available from the corresponding author upon reasonable request.

\appendix

\section{Symmetry of the potential}\label{sec:appendix}

By using the parametrization of the $\hat{x}$ and $\hat{y}$ operators by $\varphi$,the dipole-dipole interaction term of Eq.~\eqref{eq:Ham2} can be written as
\begin{align}
V=\sum_{i=1}^{N-1}&\left[(1-\frac{3}{2}\cos^2\gamma)\cos (\varphi_{i}-\varphi_{i+1})\right. \notag\\
&\left. -\frac{3}{2}\cos^2\gamma\cos (\varphi_{i}+\varphi_{i+1})\right].
\end{align}
For $\gamma=\arccos\left(\sqrt{2/3}\right)$  the potential becomes
\begin{align}
V=-\sum_{i=1}^{N-1}\cos\left(\varphi_{i}+\varphi_{i+1}\right).
\end{align}
This term is invariant with respect to a rotation of $\varphi_{i}$ by $\Delta\varphi\in\left(0,2\pi\right)$ and of its neighbor  $\varphi_{i+1}$ by $2\pi-\Delta\varphi$. For $\gamma=90^\circ$ the potential term becomes
\begin{align}
V=\sum_{i=1}^{N-1}\cos\left(\varphi_{i}-\varphi_{i+1}\right).
\end{align}
This term is invariant with respect to a rotation of all angles $\varphi$ by the same angle $\Delta\varphi\in\left(0,2\pi\right)$. 



\begin{thebibliography}{42}%
\makeatletter
\providecommand \@ifxundefined [1]{%
 \@ifx{#1\undefined}
}%
\providecommand \@ifnum [1]{%
 \ifnum #1\expandafter \@firstoftwo
 \else \expandafter \@secondoftwo
 \fi
}%
\providecommand \@ifx [1]{%
 \ifx #1\expandafter \@firstoftwo
 \else \expandafter \@secondoftwo
 \fi
}%
\providecommand \natexlab [1]{#1}%
\providecommand \enquote  [1]{``#1''}%
\providecommand \bibnamefont  [1]{#1}%
\providecommand \bibfnamefont [1]{#1}%
\providecommand \citenamefont [1]{#1}%
\providecommand \href@noop [0]{\@secondoftwo}%
\providecommand \href [0]{\begingroup \@sanitize@url \@href}%
\providecommand \@href[1]{\@@startlink{#1}\@@href}%
\providecommand \@@href[1]{\endgroup#1\@@endlink}%
\providecommand \@sanitize@url [0]{\catcode `\\12\catcode `\$12\catcode
  `\&12\catcode `\#12\catcode `\^12\catcode `\_12\catcode `\%12\relax}%
\providecommand \@@startlink[1]{}%
\providecommand \@@endlink[0]{}%
\providecommand \url  [0]{\begingroup\@sanitize@url \@url }%
\providecommand \@url [1]{\endgroup\@href {#1}{\urlprefix }}%
\providecommand \urlprefix  [0]{URL }%
\providecommand \Eprint [0]{\href }%
\providecommand \doibase [0]{http://dx.doi.org/}%
\providecommand \selectlanguage [0]{\@gobble}%
\providecommand \bibinfo  [0]{\@secondoftwo}%
\providecommand \bibfield  [0]{\@secondoftwo}%
\providecommand \translation [1]{[#1]}%
\providecommand \BibitemOpen [0]{}%
\providecommand \bibitemStop [0]{}%
\providecommand \bibitemNoStop [0]{.\EOS\space}%
\providecommand \EOS [0]{\spacefactor3000\relax}%
\providecommand \BibitemShut  [1]{\csname bibitem#1\endcsname}%
\let\auto@bib@innerbib\@empty
\bibitem [{\citenamefont {Carr}\ \emph {et~al.}(2009)\citenamefont {Carr},
  \citenamefont {DeMille}, \citenamefont {Krems},\ and\ \citenamefont
  {Ye}}]{carr2009cold}%
  \BibitemOpen
  \bibfield  {author} {\bibinfo {author} {\bibfnamefont {L.~D.}\ \bibnamefont
  {Carr}}, \bibinfo {author} {\bibfnamefont {D.}~\bibnamefont {DeMille}},
  \bibinfo {author} {\bibfnamefont {R.~V.}\ \bibnamefont {Krems}}, \ and\
  \bibinfo {author} {\bibfnamefont {J.}~\bibnamefont {Ye}},\ }\bibfield
  {title} {\enquote {\bibinfo {title} {Cold and ultracold molecules: science,
  technology and applications},}\ }\href@noop {} {\bibfield  {journal}
  {\bibinfo  {journal} {New J. Phys.}\ }\textbf {\bibinfo {volume} {11}},\
  \bibinfo {pages} {055049} (\bibinfo {year} {2009})}\BibitemShut {NoStop}%
\bibitem [{\citenamefont {Golomedov}, \citenamefont {Astrakharchik},\ and\
  \citenamefont {Lozovik}(2011)}]{golomedov2011mesoscopic}%
  \BibitemOpen
  \bibfield  {author} {\bibinfo {author} {\bibfnamefont {A.}~\bibnamefont
  {Golomedov}}, \bibinfo {author} {\bibfnamefont {G.}~\bibnamefont
  {Astrakharchik}}, \ and\ \bibinfo {author} {\bibfnamefont {Y.~E.}\
  \bibnamefont {Lozovik}},\ }\bibfield  {title} {\enquote {\bibinfo {title}
  {Mesoscopic supersolid of dipoles in a trap},}\ }\href@noop {} {\bibfield
  {journal} {\bibinfo  {journal} {Phys. Rev. A}\ }\textbf {\bibinfo {volume}
  {84}},\ \bibinfo {pages} {033615} (\bibinfo {year} {2011})}\BibitemShut
  {NoStop}%
\bibitem [{\citenamefont {Samajdar}\ \emph {et~al.}(2021)\citenamefont
  {Samajdar}, \citenamefont {Ho}, \citenamefont {Pichler}, \citenamefont
  {Lukin},\ and\ \citenamefont {Sachdev}}]{samajdar2021quantum}%
  \BibitemOpen
  \bibfield  {author} {\bibinfo {author} {\bibfnamefont {R.}~\bibnamefont
  {Samajdar}}, \bibinfo {author} {\bibfnamefont {W.~W.}\ \bibnamefont {Ho}},
  \bibinfo {author} {\bibfnamefont {H.}~\bibnamefont {Pichler}}, \bibinfo
  {author} {\bibfnamefont {M.~D.}\ \bibnamefont {Lukin}}, \ and\ \bibinfo
  {author} {\bibfnamefont {S.}~\bibnamefont {Sachdev}},\ }\bibfield  {title}
  {\enquote {\bibinfo {title} {Quantum phases of rydberg atoms on a kagome
  lattice},}\ }\href@noop {} {\bibfield  {journal} {\bibinfo  {journal} {Proc.
  Natl. Acad. Sci. U.S.A.}\ }\textbf {\bibinfo {volume} {118}},\ \bibinfo
  {pages} {e2015785118} (\bibinfo {year} {2021})}\BibitemShut {NoStop}%
\bibitem [{\citenamefont {Bao}\ \emph {et~al.}(2023)\citenamefont {Bao},
  \citenamefont {Yu}, \citenamefont {Anderegg}, \citenamefont {Chae},
  \citenamefont {Ketterle}, \citenamefont {Ni},\ and\ \citenamefont
  {Doyle}}]{baodipolar2023}%
  \BibitemOpen
  \bibfield  {author} {\bibinfo {author} {\bibfnamefont {Y.}~\bibnamefont
  {Bao}}, \bibinfo {author} {\bibfnamefont {S.~S.}\ \bibnamefont {Yu}},
  \bibinfo {author} {\bibfnamefont {L.}~\bibnamefont {Anderegg}}, \bibinfo
  {author} {\bibfnamefont {E.}~\bibnamefont {Chae}}, \bibinfo {author}
  {\bibfnamefont {W.}~\bibnamefont {Ketterle}}, \bibinfo {author}
  {\bibfnamefont {K.-K.}\ \bibnamefont {Ni}}, \ and\ \bibinfo {author}
  {\bibfnamefont {J.~M.}\ \bibnamefont {Doyle}},\ }\bibfield  {title} {\enquote
  {\bibinfo {title} {Dipolar spin-exchange and entanglement between molecules
  in an optical tweezer array},}\ }\href@noop {} {\bibfield  {journal}
  {\bibinfo  {journal} {Science}\ }\textbf {\bibinfo {volume} {382}},\ \bibinfo
  {pages} {1138} (\bibinfo {year} {2023})}\BibitemShut {NoStop}%
\bibitem [{\citenamefont {Holland}, \citenamefont {Lu},\ and\ \citenamefont
  {Cheuk}(2023)}]{connordipolar2023}%
  \BibitemOpen
  \bibfield  {author} {\bibinfo {author} {\bibfnamefont {C.~M.}\ \bibnamefont
  {Holland}}, \bibinfo {author} {\bibfnamefont {Y.}~\bibnamefont {Lu}}, \ and\
  \bibinfo {author} {\bibfnamefont {L.~W.}\ \bibnamefont {Cheuk}},\ }\bibfield
  {title} {\enquote {\bibinfo {title} {On-demand entanglement of molecules in a
  reconfigurable optical tweezer array},}\ }\href@noop {} {\bibfield  {journal}
  {\bibinfo  {journal} {Science}\ }\textbf {\bibinfo {volume} {382}},\ \bibinfo
  {pages} {1143} (\bibinfo {year} {2023})}\BibitemShut {NoStop}%
\bibitem [{\citenamefont {Vojta}(2003)}]{vojta2003quantum}%
  \BibitemOpen
  \bibfield  {author} {\bibinfo {author} {\bibfnamefont {M.}~\bibnamefont
  {Vojta}},\ }\bibfield  {title} {\enquote {\bibinfo {title} {Quantum phase
  transitions},}\ }\href@noop {} {\bibfield  {journal} {\bibinfo  {journal}
  {Rep. Prog. Phys.}\ }\textbf {\bibinfo {volume} {66}},\ \bibinfo {pages}
  {2069} (\bibinfo {year} {2003})}\BibitemShut {NoStop}%
\bibitem [{\citenamefont {Sachdev}\ and\ \citenamefont
  {Keimer}(2011)}]{sachdev2011quantumcrit}%
  \BibitemOpen
  \bibfield  {author} {\bibinfo {author} {\bibfnamefont {S.}~\bibnamefont
  {Sachdev}}\ and\ \bibinfo {author} {\bibfnamefont {B.}~\bibnamefont
  {Keimer}},\ }\bibfield  {title} {\enquote {\bibinfo {title} {Quantum
  criticality},}\ }\href@noop {} {\bibfield  {journal} {\bibinfo  {journal}
  {Phys. Today}\ }\textbf {\bibinfo {volume} {64}},\ \bibinfo {pages} {29}
  (\bibinfo {year} {2011})}\BibitemShut {NoStop}%
\bibitem [{\citenamefont {Sachdev}(2011)}]{sachdev2011quantum}%
  \BibitemOpen
  \bibfield  {author} {\bibinfo {author} {\bibfnamefont {S.}~\bibnamefont
  {Sachdev}},\ }\href@noop {} {\emph {\bibinfo {title} {Quantum phase
  transitions}}}\ (\bibinfo  {publisher} {Cambridge University Press},\
  \bibinfo {year} {2011})\BibitemShut {NoStop}%
\bibitem [{\citenamefont {Sachdev}(2000)}]{sachdev2000quantum}%
  \BibitemOpen
  \bibfield  {author} {\bibinfo {author} {\bibfnamefont {S.}~\bibnamefont
  {Sachdev}},\ }\bibfield  {title} {\enquote {\bibinfo {title} {Quantum
  criticality: competing ground states in low dimensions},}\ }\href@noop {}
  {\bibfield  {journal} {\bibinfo  {journal} {Science}\ }\textbf {\bibinfo
  {volume} {288}},\ \bibinfo {pages} {475--480} (\bibinfo {year}
  {2000})}\BibitemShut {NoStop}%
\bibitem [{\citenamefont {Anderson}, \citenamefont {Younge},\ and\
  \citenamefont {Raithel}(2011)}]{anderson2011trapping}%
  \BibitemOpen
  \bibfield  {author} {\bibinfo {author} {\bibfnamefont {S.~E.}\ \bibnamefont
  {Anderson}}, \bibinfo {author} {\bibfnamefont {K.}~\bibnamefont {Younge}}, \
  and\ \bibinfo {author} {\bibfnamefont {G.}~\bibnamefont {Raithel}},\
  }\bibfield  {title} {\enquote {\bibinfo {title} {Trapping rydberg atoms in an
  optical lattice},}\ }\href@noop {} {\bibfield  {journal} {\bibinfo  {journal}
  {Phys. Rev. Lett.}\ }\textbf {\bibinfo {volume} {107}},\ \bibinfo {pages}
  {263001} (\bibinfo {year} {2011})}\BibitemShut {NoStop}%
\bibitem [{\citenamefont {Yan}\ \emph {et~al.}(2013)\citenamefont {Yan},
  \citenamefont {Moses}, \citenamefont {Gadway}, \citenamefont {Covey},
  \citenamefont {Hazzard}, \citenamefont {Rey}, \citenamefont {Jin},\ and\
  \citenamefont {Ye}}]{yan2013observation}%
  \BibitemOpen
  \bibfield  {author} {\bibinfo {author} {\bibfnamefont {B.}~\bibnamefont
  {Yan}}, \bibinfo {author} {\bibfnamefont {S.~A.}\ \bibnamefont {Moses}},
  \bibinfo {author} {\bibfnamefont {B.}~\bibnamefont {Gadway}}, \bibinfo
  {author} {\bibfnamefont {J.~P.}\ \bibnamefont {Covey}}, \bibinfo {author}
  {\bibfnamefont {K.~R.}\ \bibnamefont {Hazzard}}, \bibinfo {author}
  {\bibfnamefont {A.~M.}\ \bibnamefont {Rey}}, \bibinfo {author} {\bibfnamefont
  {D.~S.}\ \bibnamefont {Jin}}, \ and\ \bibinfo {author} {\bibfnamefont
  {J.}~\bibnamefont {Ye}},\ }\bibfield  {title} {\enquote {\bibinfo {title}
  {Observation of dipolar spin-exchange interactions with lattice-confined
  polar molecules},}\ }\href@noop {} {\bibfield  {journal} {\bibinfo  {journal}
  {Nature}\ }\textbf {\bibinfo {volume} {501}},\ \bibinfo {pages} {521--525}
  (\bibinfo {year} {2013})}\BibitemShut {NoStop}%
\bibitem [{\citenamefont {Serwatka}\ and\ \citenamefont
  {Roy}(2022{\natexlab{a}})}]{serwatka2022ferroelectric}%
  \BibitemOpen
  \bibfield  {author} {\bibinfo {author} {\bibfnamefont {T.}~\bibnamefont
  {Serwatka}}\ and\ \bibinfo {author} {\bibfnamefont {P.-N.}\ \bibnamefont
  {Roy}},\ }\bibfield  {title} {\enquote {\bibinfo {title} {Ferroelectric water
  chains in carbon nanotubes: Creation and manipulation of ordered quantum
  phases},}\ }\href@noop {} {\bibfield  {journal} {\bibinfo  {journal} {J.
  Chem. Phys.}\ }\textbf {\bibinfo {volume} {157}},\ \bibinfo {pages} {234301}
  (\bibinfo {year} {2022}{\natexlab{a}})}\BibitemShut {NoStop}%
\bibitem [{\citenamefont {Serwatka}\ \emph {et~al.}(2023)\citenamefont
  {Serwatka}, \citenamefont {Melko}, \citenamefont {Burkov},\ and\
  \citenamefont {Roy}}]{serwatka2023quantum}%
  \BibitemOpen
  \bibfield  {author} {\bibinfo {author} {\bibfnamefont {T.}~\bibnamefont
  {Serwatka}}, \bibinfo {author} {\bibfnamefont {R.~G.}\ \bibnamefont {Melko}},
  \bibinfo {author} {\bibfnamefont {A.}~\bibnamefont {Burkov}}, \ and\ \bibinfo
  {author} {\bibfnamefont {P.-N.}\ \bibnamefont {Roy}},\ }\bibfield  {title}
  {\enquote {\bibinfo {title} {Quantum phase transition in the one-dimensional
  water chain},}\ }\href@noop {} {\bibfield  {journal} {\bibinfo  {journal}
  {Phys. Rev. Lett.}\ }\textbf {\bibinfo {volume} {130}},\ \bibinfo {pages}
  {026201} (\bibinfo {year} {2023})}\BibitemShut {NoStop}%
\bibitem [{\citenamefont {Serwatka}\ and\ \citenamefont
  {Roy}(2023)}]{serwatka2023endo}%
  \BibitemOpen
  \bibfield  {author} {\bibinfo {author} {\bibfnamefont {T.}~\bibnamefont
  {Serwatka}}\ and\ \bibinfo {author} {\bibfnamefont {P.-N.}\ \bibnamefont
  {Roy}},\ }\bibfield  {title} {\enquote {\bibinfo {title} {Quantum criticality
  and universal behavior in molecular dipolar lattices of endofullerenes},}\
  }\href@noop {} {\bibfield  {journal} {\bibinfo  {journal} {J. Phys. Chem.
  Lett.}\ }\textbf {\bibinfo {volume} {14}},\ \bibinfo {pages} {5586--5591}
  (\bibinfo {year} {2023})}\BibitemShut {NoStop}%
\bibitem [{\citenamefont {Abolins}, \citenamefont {Zillich},\ and\
  \citenamefont {Whaley}(2011)}]{abolins2011ground}%
  \BibitemOpen
  \bibfield  {author} {\bibinfo {author} {\bibfnamefont {B.}~\bibnamefont
  {Abolins}}, \bibinfo {author} {\bibfnamefont {R.}~\bibnamefont {Zillich}}, \
  and\ \bibinfo {author} {\bibfnamefont {K.}~\bibnamefont {Whaley}},\
  }\bibfield  {title} {\enquote {\bibinfo {title} {A ground state monte carlo
  approach for studies of dipolar systems with rotational degrees of
  freedom},}\ }\href@noop {} {\bibfield  {journal} {\bibinfo  {journal} {J. Low
  Temp. Phys.}\ }\textbf {\bibinfo {volume} {165}},\ \bibinfo {pages}
  {249--260} (\bibinfo {year} {2011})}\BibitemShut {NoStop}%
\bibitem [{\citenamefont {Abolins}, \citenamefont {Zillich},\ and\
  \citenamefont {Whaley}(2013)}]{abolins2013erratum}%
  \BibitemOpen
  \bibfield  {author} {\bibinfo {author} {\bibfnamefont {B.}~\bibnamefont
  {Abolins}}, \bibinfo {author} {\bibfnamefont {R.}~\bibnamefont {Zillich}}, \
  and\ \bibinfo {author} {\bibfnamefont {K.}~\bibnamefont {Whaley}},\
  }\bibfield  {title} {\enquote {\bibinfo {title} {Erratum to: A ground state
  monte carlo approach for studies of dipolar systems with rotational degrees
  of freedom},}\ }\href@noop {} {\bibfield  {journal} {\bibinfo  {journal} {J.
  Low Temp. Phys.}\ }\textbf {\bibinfo {volume} {170}},\ \bibinfo {pages}
  {131--131} (\bibinfo {year} {2013})}\BibitemShut {NoStop}%
\bibitem [{\citenamefont {Abolins}, \citenamefont {Zillich},\ and\
  \citenamefont {Whaley}(2018)}]{abolins2018quantum}%
  \BibitemOpen
  \bibfield  {author} {\bibinfo {author} {\bibfnamefont {B.~P.}\ \bibnamefont
  {Abolins}}, \bibinfo {author} {\bibfnamefont {R.~E.}\ \bibnamefont
  {Zillich}}, \ and\ \bibinfo {author} {\bibfnamefont {K.~B.}\ \bibnamefont
  {Whaley}},\ }\bibfield  {title} {\enquote {\bibinfo {title} {Quantum phases
  of dipolar rotors on two-dimensional lattices},}\ }\href@noop {} {\bibfield
  {journal} {\bibinfo  {journal} {J. Chem. Phys.}\ }\textbf {\bibinfo {volume}
  {148}} (\bibinfo {year} {2018})}\BibitemShut {NoStop}%
\bibitem [{\citenamefont {Gorshunov}\ \emph {et~al.}(2016)\citenamefont
  {Gorshunov}, \citenamefont {Torgashev}, \citenamefont {Zhukova},
  \citenamefont {Thomas}, \citenamefont {Belyanchikov}, \citenamefont {Kadlec},
  \citenamefont {Kadlec}, \citenamefont {Savinov}, \citenamefont {Ostapchuk},
  \citenamefont {Petzelt} \emph {et~al.}}]{gorshunov2016incipient}%
  \BibitemOpen
  \bibfield  {author} {\bibinfo {author} {\bibfnamefont {B.}~\bibnamefont
  {Gorshunov}}, \bibinfo {author} {\bibfnamefont {V.}~\bibnamefont
  {Torgashev}}, \bibinfo {author} {\bibfnamefont {E.}~\bibnamefont {Zhukova}},
  \bibinfo {author} {\bibfnamefont {V.}~\bibnamefont {Thomas}}, \bibinfo
  {author} {\bibfnamefont {M.}~\bibnamefont {Belyanchikov}}, \bibinfo {author}
  {\bibfnamefont {C.}~\bibnamefont {Kadlec}}, \bibinfo {author} {\bibfnamefont
  {F.}~\bibnamefont {Kadlec}}, \bibinfo {author} {\bibfnamefont
  {M.}~\bibnamefont {Savinov}}, \bibinfo {author} {\bibfnamefont
  {T.}~\bibnamefont {Ostapchuk}}, \bibinfo {author} {\bibfnamefont
  {J.}~\bibnamefont {Petzelt}},  \emph {et~al.},\ }\bibfield  {title} {\enquote
  {\bibinfo {title} {Incipient ferroelectricity of water molecules confined to
  nano-channels of beryl},}\ }\href@noop {} {\bibfield  {journal} {\bibinfo
  {journal} {Nat. Commun.}\ }\textbf {\bibinfo {volume} {7}},\ \bibinfo {pages}
  {1--10} (\bibinfo {year} {2016})}\BibitemShut {NoStop}%
\bibitem [{\citenamefont {Belyanchikov}\ \emph {et~al.}(2020)\citenamefont
  {Belyanchikov}, \citenamefont {Savinov}, \citenamefont {Bedran},
  \citenamefont {Bednyakov}, \citenamefont {Proschek}, \citenamefont
  {Prokleska}, \citenamefont {Abalmasov}, \citenamefont {Petzelt},
  \citenamefont {Zhukova}, \citenamefont {Thomas} \emph
  {et~al.}}]{belyanchikov2020dielectric}%
  \BibitemOpen
  \bibfield  {author} {\bibinfo {author} {\bibfnamefont {M.}~\bibnamefont
  {Belyanchikov}}, \bibinfo {author} {\bibfnamefont {M.}~\bibnamefont
  {Savinov}}, \bibinfo {author} {\bibfnamefont {Z.}~\bibnamefont {Bedran}},
  \bibinfo {author} {\bibfnamefont {P.}~\bibnamefont {Bednyakov}}, \bibinfo
  {author} {\bibfnamefont {P.}~\bibnamefont {Proschek}}, \bibinfo {author}
  {\bibfnamefont {J.}~\bibnamefont {Prokleska}}, \bibinfo {author}
  {\bibfnamefont {V.}~\bibnamefont {Abalmasov}}, \bibinfo {author}
  {\bibfnamefont {J.}~\bibnamefont {Petzelt}}, \bibinfo {author} {\bibfnamefont
  {E.}~\bibnamefont {Zhukova}}, \bibinfo {author} {\bibfnamefont
  {V.}~\bibnamefont {Thomas}},  \emph {et~al.},\ }\bibfield  {title} {\enquote
  {\bibinfo {title} {Dielectric ordering of water molecules arranged in a
  dipolar lattice},}\ }\href@noop {} {\bibfield  {journal} {\bibinfo  {journal}
  {Nat. Commun.}\ }\textbf {\bibinfo {volume} {11}},\ \bibinfo {pages} {1--9}
  (\bibinfo {year} {2020})}\BibitemShut {NoStop}%
\bibitem [{\citenamefont {Kolesnikov}\ \emph {et~al.}(2016)\citenamefont
  {Kolesnikov}, \citenamefont {Reiter}, \citenamefont {Choudhury},
  \citenamefont {Prisk}, \citenamefont {Mamontov}, \citenamefont {Podlesnyak},
  \citenamefont {Ehlers}, \citenamefont {Seel}, \citenamefont {Wesolowski},\
  and\ \citenamefont {Anovitz}}]{kolesnikov2016quantum}%
  \BibitemOpen
  \bibfield  {author} {\bibinfo {author} {\bibfnamefont {A.~I.}\ \bibnamefont
  {Kolesnikov}}, \bibinfo {author} {\bibfnamefont {G.~F.}\ \bibnamefont
  {Reiter}}, \bibinfo {author} {\bibfnamefont {N.}~\bibnamefont {Choudhury}},
  \bibinfo {author} {\bibfnamefont {T.~R.}\ \bibnamefont {Prisk}}, \bibinfo
  {author} {\bibfnamefont {E.}~\bibnamefont {Mamontov}}, \bibinfo {author}
  {\bibfnamefont {A.}~\bibnamefont {Podlesnyak}}, \bibinfo {author}
  {\bibfnamefont {G.}~\bibnamefont {Ehlers}}, \bibinfo {author} {\bibfnamefont
  {A.~G.}\ \bibnamefont {Seel}}, \bibinfo {author} {\bibfnamefont {D.~J.}\
  \bibnamefont {Wesolowski}}, \ and\ \bibinfo {author} {\bibfnamefont {L.~M.}\
  \bibnamefont {Anovitz}},\ }\bibfield  {title} {\enquote {\bibinfo {title}
  {Quantum tunneling of water in beryl: a new state of the water molecule},}\
  }\href@noop {} {\bibfield  {journal} {\bibinfo  {journal} {Phys. Rev. Lett.}\
  }\textbf {\bibinfo {volume} {116}},\ \bibinfo {pages} {167802} (\bibinfo
  {year} {2016})}\BibitemShut {NoStop}%
\bibitem [{\citenamefont {Belyanchikov}\ \emph {et~al.}(2022)\citenamefont
  {Belyanchikov}, \citenamefont {Bedran}, \citenamefont {Savinov},
  \citenamefont {Bednyakov}, \citenamefont {Proschek}, \citenamefont
  {Prokleska}, \citenamefont {Abalmasov}, \citenamefont {Zhukova},
  \citenamefont {Thomas}, \citenamefont {Dudka} \emph
  {et~al.}}]{belyanchikov2022single}%
  \BibitemOpen
  \bibfield  {author} {\bibinfo {author} {\bibfnamefont {M.}~\bibnamefont
  {Belyanchikov}}, \bibinfo {author} {\bibfnamefont {Z.}~\bibnamefont
  {Bedran}}, \bibinfo {author} {\bibfnamefont {M.}~\bibnamefont {Savinov}},
  \bibinfo {author} {\bibfnamefont {P.}~\bibnamefont {Bednyakov}}, \bibinfo
  {author} {\bibfnamefont {P.}~\bibnamefont {Proschek}}, \bibinfo {author}
  {\bibfnamefont {J.}~\bibnamefont {Prokleska}}, \bibinfo {author}
  {\bibfnamefont {V.}~\bibnamefont {Abalmasov}}, \bibinfo {author}
  {\bibfnamefont {E.}~\bibnamefont {Zhukova}}, \bibinfo {author} {\bibfnamefont
  {V.}~\bibnamefont {Thomas}}, \bibinfo {author} {\bibfnamefont
  {A.}~\bibnamefont {Dudka}},  \emph {et~al.},\ }\bibfield  {title} {\enquote
  {\bibinfo {title} {Single-particle and collective excitations of polar water
  molecules confined in nano-pores within a cordierite crystal lattice},}\
  }\href@noop {} {\bibfield  {journal} {\bibinfo  {journal} {Phys. Chem. Chem.
  Phys.}\ }\textbf {\bibinfo {volume} {24}},\ \bibinfo {pages} {6890--6904}
  (\bibinfo {year} {2022})}\BibitemShut {NoStop}%
\bibitem [{\citenamefont {Felker}\ and\ \citenamefont
  {Ba{\v{c}}i{\'c}}(2017{\natexlab{a}})}]{felker2017electric}%
  \BibitemOpen
  \bibfield  {author} {\bibinfo {author} {\bibfnamefont {P.~M.}\ \bibnamefont
  {Felker}}\ and\ \bibinfo {author} {\bibfnamefont {Z.}~\bibnamefont
  {Ba{\v{c}}i{\'c}}},\ }\bibfield  {title} {\enquote {\bibinfo {title}
  {Electric-dipole-coupled h2o@ c60 dimer: Translation-rotation eigenstates
  from twelve-dimensional quantum calculations},}\ }\href@noop {} {\bibfield
  {journal} {\bibinfo  {journal} {J. Chem. Phys.}\ }\textbf {\bibinfo {volume}
  {146}},\ \bibinfo {pages} {084303} (\bibinfo {year}
  {2017}{\natexlab{a}})}\BibitemShut {NoStop}%
\bibitem [{\citenamefont {Felker}\ and\ \citenamefont
  {Ba{\v{c}}i{\'c}}(2017{\natexlab{b}})}]{felker2017accurate}%
  \BibitemOpen
  \bibfield  {author} {\bibinfo {author} {\bibfnamefont {P.~M.}\ \bibnamefont
  {Felker}}\ and\ \bibinfo {author} {\bibfnamefont {Z.}~\bibnamefont
  {Ba{\v{c}}i{\'c}}},\ }\bibfield  {title} {\enquote {\bibinfo {title}
  {Accurate quantum calculations of translation-rotation eigenstates in
  electric-dipole-coupled h2o@ c60 assemblies},}\ }\href@noop {} {\bibfield
  {journal} {\bibinfo  {journal} {Chem. Phys. Lett.}\ }\textbf {\bibinfo
  {volume} {683}},\ \bibinfo {pages} {172--178} (\bibinfo {year}
  {2017}{\natexlab{b}})}\BibitemShut {NoStop}%
\bibitem [{\citenamefont {Halverson}, \citenamefont {Iouchtchenko},\ and\
  \citenamefont {Roy}(2018)}]{halverson2018quantifying}%
  \BibitemOpen
  \bibfield  {author} {\bibinfo {author} {\bibfnamefont {T.}~\bibnamefont
  {Halverson}}, \bibinfo {author} {\bibfnamefont {D.}~\bibnamefont
  {Iouchtchenko}}, \ and\ \bibinfo {author} {\bibfnamefont {P.-N.}\
  \bibnamefont {Roy}},\ }\bibfield  {title} {\enquote {\bibinfo {title}
  {Quantifying entanglement of rotor chains using basis truncation: Application
  to dipolar endofullerene peapods},}\ }\href@noop {} {\bibfield  {journal}
  {\bibinfo  {journal} {J. Chem. Phys.}\ }\textbf {\bibinfo {volume} {148}},\
  \bibinfo {pages} {074112} (\bibinfo {year} {2018})}\BibitemShut {NoStop}%
\bibitem [{\citenamefont {Sahoo}\ \emph {et~al.}(2020)\citenamefont {Sahoo},
  \citenamefont {Iouchtchenko}, \citenamefont {Herdman},\ and\ \citenamefont
  {Roy}}]{sahoo2020path}%
  \BibitemOpen
  \bibfield  {author} {\bibinfo {author} {\bibfnamefont {T.}~\bibnamefont
  {Sahoo}}, \bibinfo {author} {\bibfnamefont {D.}~\bibnamefont {Iouchtchenko}},
  \bibinfo {author} {\bibfnamefont {C.}~\bibnamefont {Herdman}}, \ and\
  \bibinfo {author} {\bibfnamefont {P.-N.}\ \bibnamefont {Roy}},\ }\bibfield
  {title} {\enquote {\bibinfo {title} {A path integral ground state replica
  trick approach for the computation of entanglement entropy of dipolar linear
  rotors},}\ }\href@noop {} {\bibfield  {journal} {\bibinfo  {journal} {J.
  Chem. Phys.}\ }\textbf {\bibinfo {volume} {152}},\ \bibinfo {pages} {184113}
  (\bibinfo {year} {2020})}\BibitemShut {NoStop}%
\bibitem [{\citenamefont {Sahoo}\ and\ \citenamefont
  {Gangopadhyay}(2023)}]{sahoo2023effect}%
  \BibitemOpen
  \bibfield  {author} {\bibinfo {author} {\bibfnamefont {T.}~\bibnamefont
  {Sahoo}}\ and\ \bibinfo {author} {\bibfnamefont {G.}~\bibnamefont
  {Gangopadhyay}},\ }\bibfield  {title} {\enquote {\bibinfo {title} {Effect of
  neighbouring molecules on ground-state properties of many-body polar linear
  rotor systems},}\ }\href@noop {} {\bibfield  {journal} {\bibinfo  {journal}
  {Mol. Phys.}\ ,\ \bibinfo {pages} {e2242967}} (\bibinfo {year}
  {2023})}\BibitemShut {NoStop}%
\bibitem [{\citenamefont {Sahoo}, \citenamefont {Serwatka},\ and\ \citenamefont
  {Roy}(2021)}]{sahoo2021path}%
  \BibitemOpen
  \bibfield  {author} {\bibinfo {author} {\bibfnamefont {T.}~\bibnamefont
  {Sahoo}}, \bibinfo {author} {\bibfnamefont {T.}~\bibnamefont {Serwatka}}, \
  and\ \bibinfo {author} {\bibfnamefont {P.-N.}\ \bibnamefont {Roy}},\
  }\bibfield  {title} {\enquote {\bibinfo {title} {A path integral ground state
  approach for asymmetric top rotors with nuclear spin symmetry: Application to
  water chains},}\ }\href@noop {} {\bibfield  {journal} {\bibinfo  {journal}
  {J. Chem. Phys.}\ }\textbf {\bibinfo {volume} {154}},\ \bibinfo {pages}
  {244305} (\bibinfo {year} {2021})}\BibitemShut {NoStop}%
\bibitem [{\citenamefont {White}(1992)}]{white1992density}%
  \BibitemOpen
  \bibfield  {author} {\bibinfo {author} {\bibfnamefont {S.~R.}\ \bibnamefont
  {White}},\ }\bibfield  {title} {\enquote {\bibinfo {title} {Density matrix
  formulation for quantum renormalization groups},}\ }\href@noop {} {\bibfield
  {journal} {\bibinfo  {journal} {Phys. Rev. Lett}\ }\textbf {\bibinfo {volume}
  {69}},\ \bibinfo {pages} {2863} (\bibinfo {year} {1992})}\BibitemShut
  {NoStop}%
\bibitem [{\citenamefont {Iouchtchenko}\ and\ \citenamefont
  {Roy}(2018)}]{iouchtchenko2018ground}%
  \BibitemOpen
  \bibfield  {author} {\bibinfo {author} {\bibfnamefont {D.}~\bibnamefont
  {Iouchtchenko}}\ and\ \bibinfo {author} {\bibfnamefont {P.-N.}\ \bibnamefont
  {Roy}},\ }\bibfield  {title} {\enquote {\bibinfo {title} {Ground states of
  linear rotor chains via the density matrix renormalization group},}\
  }\href@noop {} {\bibfield  {journal} {\bibinfo  {journal} {The Journal of
  Chemical Physics}\ }\textbf {\bibinfo {volume} {148}},\ \bibinfo {pages}
  {134115} (\bibinfo {year} {2018})}\BibitemShut {NoStop}%
\bibitem [{\citenamefont {Mainali}\ \emph {et~al.}(2021)\citenamefont
  {Mainali}, \citenamefont {Gatti}, \citenamefont {Iouchtchenko}, \citenamefont
  {Roy},\ and\ \citenamefont {Meyer}}]{mainali2021comparison}%
  \BibitemOpen
  \bibfield  {author} {\bibinfo {author} {\bibfnamefont {S.}~\bibnamefont
  {Mainali}}, \bibinfo {author} {\bibfnamefont {F.}~\bibnamefont {Gatti}},
  \bibinfo {author} {\bibfnamefont {D.}~\bibnamefont {Iouchtchenko}}, \bibinfo
  {author} {\bibfnamefont {P.-N.}\ \bibnamefont {Roy}}, \ and\ \bibinfo
  {author} {\bibfnamefont {H.-D.}\ \bibnamefont {Meyer}},\ }\bibfield  {title}
  {\enquote {\bibinfo {title} {Comparison of the multi-layer
  multi-configuration time-dependent hartree (ml-mctdh) method and the density
  matrix renormalization group (dmrg) for ground state properties of linear
  rotor chains},}\ }\href@noop {} {\bibfield  {journal} {\bibinfo  {journal}
  {The Journal of Chemical Physics}\ }\textbf {\bibinfo {volume} {154}},\
  \bibinfo {pages} {174106} (\bibinfo {year} {2021})}\BibitemShut {NoStop}%
\bibitem [{\citenamefont {Serwatka}\ and\ \citenamefont
  {Roy}(2022{\natexlab{b}})}]{serwatka2022ground}%
  \BibitemOpen
  \bibfield  {author} {\bibinfo {author} {\bibfnamefont {T.}~\bibnamefont
  {Serwatka}}\ and\ \bibinfo {author} {\bibfnamefont {P.-N.}\ \bibnamefont
  {Roy}},\ }\bibfield  {title} {\enquote {\bibinfo {title} {Ground state of
  asymmetric tops with dmrg: Water in one dimension},}\ }\href@noop {}
  {\bibfield  {journal} {\bibinfo  {journal} {J. Chem. Phys.}\ }\textbf
  {\bibinfo {volume} {156}},\ \bibinfo {pages} {044116} (\bibinfo {year}
  {2022}{\natexlab{b}})}\BibitemShut {NoStop}%
\bibitem [{\citenamefont {Schollw{\"o}ck}(2011)}]{schollwock2011density}%
  \BibitemOpen
  \bibfield  {author} {\bibinfo {author} {\bibfnamefont {U.}~\bibnamefont
  {Schollw{\"o}ck}},\ }\bibfield  {title} {\enquote {\bibinfo {title} {The
  density-matrix renormalization group in the age of matrix product states},}\
  }\href@noop {} {\bibfield  {journal} {\bibinfo  {journal} {Ann. Phys.}\
  }\textbf {\bibinfo {volume} {326}},\ \bibinfo {pages} {96--192} (\bibinfo
  {year} {2011})}\BibitemShut {NoStop}%
\bibitem [{\citenamefont {Fishman}, \citenamefont {White},\ and\ \citenamefont
  {Stoudenmire}(2022)}]{fishman2022itensor}%
  \BibitemOpen
  \bibfield  {author} {\bibinfo {author} {\bibfnamefont {M.}~\bibnamefont
  {Fishman}}, \bibinfo {author} {\bibfnamefont {S.}~\bibnamefont {White}}, \
  and\ \bibinfo {author} {\bibfnamefont {E.}~\bibnamefont {Stoudenmire}},\
  }\bibfield  {title} {\enquote {\bibinfo {title} {The itensor software library
  for tensor network calculations},}\ }\href@noop {} {\bibfield  {journal}
  {\bibinfo  {journal} {SciPost Physics Codebases}\ ,\ \bibinfo {pages} {004}}
  (\bibinfo {year} {2022})}\BibitemShut {NoStop}%
\bibitem [{\citenamefont {Katsura}(2012)}]{katsura2012sine}%
  \BibitemOpen
  \bibfield  {author} {\bibinfo {author} {\bibfnamefont {H.}~\bibnamefont
  {Katsura}},\ }\bibfield  {title} {\enquote {\bibinfo {title} {Sine-square
  deformation of solvable spin chains and conformal field theories},}\
  }\href@noop {} {\bibfield  {journal} {\bibinfo  {journal} {J. Phys. A: Math.
  Theor.}\ }\textbf {\bibinfo {volume} {45}},\ \bibinfo {pages} {115003}
  (\bibinfo {year} {2012})}\BibitemShut {NoStop}%
\bibitem [{\citenamefont {Hotta}\ and\ \citenamefont
  {Shibata}(2012)}]{hotta2012grand}%
  \BibitemOpen
  \bibfield  {author} {\bibinfo {author} {\bibfnamefont {C.}~\bibnamefont
  {Hotta}}\ and\ \bibinfo {author} {\bibfnamefont {N.}~\bibnamefont
  {Shibata}},\ }\bibfield  {title} {\enquote {\bibinfo {title} {Grand canonical
  finite-size numerical approaches: A route to measuring bulk properties in an
  applied field},}\ }\href@noop {} {\bibfield  {journal} {\bibinfo  {journal}
  {Phys. Rev. B}\ }\textbf {\bibinfo {volume} {86}},\ \bibinfo {pages} {041108}
  (\bibinfo {year} {2012})}\BibitemShut {NoStop}%
\bibitem [{\citenamefont {Binder}(1981{\natexlab{a}})}]{binder1981finite}%
  \BibitemOpen
  \bibfield  {author} {\bibinfo {author} {\bibfnamefont {K.}~\bibnamefont
  {Binder}},\ }\bibfield  {title} {\enquote {\bibinfo {title} {Finite size
  scaling analysis of ising model block distribution functions},}\ }\href@noop
  {} {\bibfield  {journal} {\bibinfo  {journal} {Z. Phys. B}\ }\textbf
  {\bibinfo {volume} {43}},\ \bibinfo {pages} {119--140} (\bibinfo {year}
  {1981}{\natexlab{a}})}\BibitemShut {NoStop}%
\bibitem [{\citenamefont {Binder}(1981{\natexlab{b}})}]{binder1981critical}%
  \BibitemOpen
  \bibfield  {author} {\bibinfo {author} {\bibfnamefont {K.}~\bibnamefont
  {Binder}},\ }\bibfield  {title} {\enquote {\bibinfo {title} {Critical
  properties from monte carlo coarse graining and renormalization},}\
  }\href@noop {} {\bibfield  {journal} {\bibinfo  {journal} {Phys. Rev. Lett.}\
  }\textbf {\bibinfo {volume} {47}},\ \bibinfo {pages} {693} (\bibinfo {year}
  {1981}{\natexlab{b}})}\BibitemShut {NoStop}%
\bibitem [{\citenamefont {Calabrese}\ and\ \citenamefont
  {Cardy}(2009)}]{calabrese2009entanglement}%
  \BibitemOpen
  \bibfield  {author} {\bibinfo {author} {\bibfnamefont {P.}~\bibnamefont
  {Calabrese}}\ and\ \bibinfo {author} {\bibfnamefont {J.}~\bibnamefont
  {Cardy}},\ }\bibfield  {title} {\enquote {\bibinfo {title} {Entanglement
  entropy and conformal field theory},}\ }\href@noop {} {\bibfield  {journal}
  {\bibinfo  {journal} {J. Phys. A: math. Theor.}\ }\textbf {\bibinfo {volume}
  {42}},\ \bibinfo {pages} {504005} (\bibinfo {year} {2009})}\BibitemShut
  {NoStop}%
\bibitem [{\citenamefont {Calabrese}\ and\ \citenamefont
  {Cardy}(2004)}]{calabrese2004entanglement}%
  \BibitemOpen
  \bibfield  {author} {\bibinfo {author} {\bibfnamefont {P.}~\bibnamefont
  {Calabrese}}\ and\ \bibinfo {author} {\bibfnamefont {J.}~\bibnamefont
  {Cardy}},\ }\bibfield  {title} {\enquote {\bibinfo {title} {Entanglement
  entropy and quantum field theory},}\ }\href@noop {} {\bibfield  {journal}
  {\bibinfo  {journal} {J. Stat. Mech.: Theor. Exp.}\ }\textbf {\bibinfo
  {volume} {2004}},\ \bibinfo {pages} {P06002} (\bibinfo {year}
  {2004})}\BibitemShut {NoStop}%
\bibitem [{\citenamefont {Eisert}, \citenamefont {Cramer},\ and\ \citenamefont
  {Plenio}(2010)}]{eisert2010colloquium}%
  \BibitemOpen
  \bibfield  {author} {\bibinfo {author} {\bibfnamefont {J.}~\bibnamefont
  {Eisert}}, \bibinfo {author} {\bibfnamefont {M.}~\bibnamefont {Cramer}}, \
  and\ \bibinfo {author} {\bibfnamefont {M.~B.}\ \bibnamefont {Plenio}},\
  }\bibfield  {title} {\enquote {\bibinfo {title} {Colloquium: Area laws for
  the entanglement entropy},}\ }\href@noop {} {\bibfield  {journal} {\bibinfo
  {journal} {Rev. Mod. Phys.}\ }\textbf {\bibinfo {volume} {82}},\ \bibinfo
  {pages} {277} (\bibinfo {year} {2010})}\BibitemShut {NoStop}%
\bibitem [{\citenamefont {Ibarra-Garc{\'\i}a-Padilla}, \citenamefont
  {Malanche-Flores},\ and\ \citenamefont
  {Poveda-Cuevas}(2016)}]{ibarra2016hobbyhorse}%
  \BibitemOpen
  \bibfield  {author} {\bibinfo {author} {\bibfnamefont {E.}~\bibnamefont
  {Ibarra-Garc{\'\i}a-Padilla}}, \bibinfo {author} {\bibfnamefont {C.~G.}\
  \bibnamefont {Malanche-Flores}}, \ and\ \bibinfo {author} {\bibfnamefont
  {F.~J.}\ \bibnamefont {Poveda-Cuevas}},\ }\bibfield  {title} {\enquote
  {\bibinfo {title} {The hobbyhorse of magnetic systems: the ising model},}\
  }\href@noop {} {\bibfield  {journal} {\bibinfo  {journal} {Eur. J. Phys.}\
  }\textbf {\bibinfo {volume} {37}},\ \bibinfo {pages} {065103} (\bibinfo
  {year} {2016})}\BibitemShut {NoStop}%
\bibitem [{\citenamefont {Kosterlitz}(1974)}]{kosterlitz1974critical}%
  \BibitemOpen
  \bibfield  {author} {\bibinfo {author} {\bibfnamefont {J.}~\bibnamefont
  {Kosterlitz}},\ }\bibfield  {title} {\enquote {\bibinfo {title} {The critical
  properties of the two-dimensional xy model},}\ }\href@noop {} {\bibfield
  {journal} {\bibinfo  {journal} {J. Phys. C: Solid State Phys.}\ }\textbf
  {\bibinfo {volume} {7}},\ \bibinfo {pages} {1046} (\bibinfo {year}
  {1974})}\BibitemShut {NoStop}%
\end{thebibliography}
%

\end{document}